\documentclass[article]{aa}  
\usepackage{natbib}
\bibpunct{(}{)}{;}{a}{}{,}
\usepackage{graphicx}
\usepackage{color}
\usepackage{url}
\usepackage[colorlinks=true,linkcolor=blue,citecolor=blue]{hyperref} 
\usepackage{float}
\usepackage{orcidlink}
\usepackage{soul}
\usepackage{arydshln}
\usepackage[utf8]{inputenc}
\usepackage{pifont} %

\usepackage[normalem]{ulem}
\usepackage{xspace}
\usepackage{txfonts}

\newcommand{\psrtar}{MAXI\,J1957+032}

\newcommand{\gtap}{\mathrel{\hbox{\rlap{\lower.55ex \hbox {$\sim$}}
                   \kern-.3em \raise.4ex \hbox{$>$}}}}
\newcommand{\ltap}{\mathrel{\hbox{\rlap{\lower.55ex \hbox {$\sim$}}
                   \kern-.3em \raise.4ex \hbox{$<$}}}}

\newcommand{\nustar}{{NuSTAR}\xspace}
\newcommand{\ep}{{EP}\xspace}
\newcommand{\epfull}{{Einstein Probe}\xspace}
\newcommand{\nicer}{{NICER}\xspace} %

\newcommand{\xmm}{{XMM-Newton}\xspace}
\newcommand{\swift}{{\it Swift}\xspace}
\newcommand{\maxi}{{MAXI}\xspace} %

\newcommand{\hxmt}{{{\it Insight}-HXMT}\xspace}

\def\be{\begin{equation}} 
\def\ee{\end{equation}}

\begin{document} 

   \title{X-ray and radio observations of the AMXP MAXI J1957+032 covering the 2022--2025 outbursts}
   \titlerunning{X-ray and radio observations of \psrtar}
   \authorrunning{Z. Li et al.}
   \author{Zhaosheng Li\inst{1}
           \and
              Lucien Kuiper\inst{2}
            \and
              Yuanyue Pan\inst{1}
    \and
    Renxin Xu\inst{3,4}
    \and
    Mingyu Ge\inst{5}
    \and
    Shanshan Weng\inst{6}
    \and
    Long Peng\inst{7}
    \and
    Wenhui Yu\inst{7}
    \and
    Yue Huang\inst{5}
    \and
    Liang Zhang\inst{5}
    \and
    Liming Song\inst{5,8}
    \and
    Sergey V. Molkov\inst{9}
    \and
    Alexander A. Lutovinov\inst{9}
    \and
    Shu Zhang\inst{5}
    \and
    Shuang-Nan Zhang\inst{5}
          }
   \offprints{Z. Li}

   \institute{School of Science, Qingdao University of Technology, Qingdao 266525, P.R. China\\
              \email{lizhaosheng@xtu.edu.cn}
               \and
               SRON - Space Research Organisation Netherlands, Niels Bohrweg 4, 2333 CA, Leiden, The Netherlands
              \and
              Department of Astronomy, School of Physics, Peking University, Beijing 100871, People's Republic of China
              \and
              Kavli Institute for Astronomy and Astrophysics, Peking University, Beijing 100871, People's Republic of China
              \and
              Key Laboratory of Particle Astrophysics, Institute of High Energy Physics, Chinese Academy of Sciences, 19B Yuquan Road, Beijing 100049, China
              \and
              Department of Physics and Institute of Theoretical Physics, Nanjing Normal University, Nanjing 210023, People’s Republic of China
              \and
              Key Laboratory of Stars and Interstellar Medium, Xiangtan University, Xiangtan 411105, Hunan, P.R. China
            \and
              University of Chinese Academy of Sciences, Beijing, 100049, China
              \and
              Space Research Institute, Russian Academy of Sciences, Profsoyuznaya 84/32, 117997 Moscow, Russia
              }

   \date{Received xx  / Accepted xx}

  \abstract{
We presented a comprehensive multi-epoch timing and multiwavelength analysis of the accreting millisecond X-ray pulsar \psrtar, covering two major outbursts in 2022 and 2025.  By reanalyzing the 2022 outburst data from the \textit{Neutron Star Interior Composition Explorer} (\nicer), we found the spin frequency and orbital parameters from the observations in 0.3--5 keV. For the 2025 outburst, we reported the detection of pulsations with the \textit{Einstein Probe} (EP). Based on the $\sim$3-year baseline between these two outbursts, we measured a significant long-term spin-down rate of $\dot\nu = (-5.73 \pm 0.28) \times 10^{-14}~{\rm Hz~s^{-1}}$. Assuming that the quiescent spin-down is driven by magnetic dipole radiation, we inferred a spin-down luminosity of $L \approx 1.1 \times 10^{36}~{\rm erg~s^{-1}}$ and a surface dipolar magnetic field of $B \approx (7.3 - 10.4) \times 10^8$~G. Furthermore, we conducted a deep radio pulsation search with the \textit{Five-hundred-meter Aperture Spherical radio Telescope} (FAST) during the X-ray quiescent state in 2024, resulting in a non-detection with a 7$\sigma$ flux density upper limit of 12.3 $\mu$Jy. This corresponds to a radio efficiency upper limit of $\xi < 2.8 \times 10^{-10}$, which is significantly lower than that of typical millisecond pulsars with a similar spin-down power. This profound radio pulsation faintness can be explained by two primary scenarios: either a geometric effect, wherein the pulsar's radio beam is directed away from our line of sight, or a physical suppression of the emission mechanism, potentially caused by a persistent low-level accretion flow during the X-ray quiescent state.}


   \maketitle

\section{Introduction}
The transient X-ray source \psrtar\ was first discovered on May 11, 2015, by the \textit{Monitor of All-sky X-ray Image} (\maxi), which detected a faint and short-lived outburst \citep{ATel7504}. Coincident with an independent discovery by \textit{INTEGRAL} \citep{ATel7506}, rapid follow-up observations with the \textit{Neil Gehrels Swift Observatory} and the GROND optical telescope quickly localized the source and confirmed its nature as a fast transient, with the initial outburst decaying on a timescale of less than a day \citep{ATel7520, ATel7524}.

Since its discovery, \psrtar\ has established itself as a prolific recurrent transient, and its activity has become well characterized. The source typically underwent outbursts that were consistently short, lasting only a few days, and relatively dim, with peak luminosities of a few tens of milli-Crab. This pattern was observed in subsequent outbursts in October 2015 \citep{ATel8143} and January 2016 \citep{ATel8529}. A notable exception occurred in September 2016, when the source exhibited an outburst that was approximately 20 times brighter and spectrally harder than previously seen \citep{ATel9572}. An observation of this bright outburst with \textit{Chandra} provided the sub-arcsecond localization for the source, which confirmed the association with the optical counterpart and revealed significant spectral evolution during the outburst decay \citep{ATel9591}.  A comprehensive analysis of these first four outbursts was presented by \citet{MataSanchez2017}. They highlighted that the combination of short outburst durations, frequent recurrence, and a featureless blue optical spectrum was strongly reminiscent of the known accreting millisecond X-ray pulsar (AMXP) population. Based on these characteristics, they were the first to propose that \psrtar\ was an AMXP candidate. Furthermore, by assuming a peak outburst luminosity typical for AMXPs ($\sim$1\% of the Eddington limit), they estimated a source distance of approximately 5--6 kpc, which was refined to $5\pm2$ kpc by \citet{Ravi17}.

This AMXP prediction was definitively confirmed during the source's 2022 outburst \citep{ATel15440}, when the \textit{Neutron Star Interior Composition Explorer} (\nicer) discovered coherent pulsations at $\sim$314~Hz. This detection firmly identified \psrtar\ as an AMXP in a ~1-hour ultracompact binary orbit \citep{Sanna22}.  Radio observations during this outburst yielded a non-detection, suggesting that the source was radio-quiet while active in X-rays \citep{ATel15462}. Most recently, a new outburst in May 2025 was discovered independently by \epfull(EP)/WXT  and \maxi \citep{GCN40375,ATel17170}. Highlighting its unique rapid follow-up capabilities, \ep\ detected the $\sim$314~Hz pulsations during this outburst  \citep{2025ATel17279}. This crucial detection provided a second, independent timing solution separated by a baseline of approximately three years.

In this work we report on the analysis of \psrtar\ data from \nicer\ observations performed during its 2022 and 2025 outbursts, and from \ep, \hxmt, and \nustar\ observations during its 2025 outburst, as well as from FAST radio observations during the X-ray quiescent state between these two outbursts. In Sect.~\ref{sec:obs}, we introduce the observations and data analysis. The timing and spectral results are reported in Sects.~\ref{sec:timing} and \ref{sec:spec}, respectively. We discuss our main findings in Sect.~\ref{sec:dis}.
\section{X-ray and radio observations}\label{sec:obs}

\begin{table}%
{\small
\caption{X-ray observations of \psrtar\ for the 2022 and 2025 outbursts. }\label{table:x-ray-observations} 
\centering
\begin{tabular}{lccc} 
\hline 
Mission & Obs. ID & Instrument  & Exposure \\
 &  &  &(ks) \\
\hline 
\noalign{\smallskip}  
\multicolumn{4}{c}{2022 outburst} \\
\noalign{\smallskip}  

\hline 
\nicer        &  5202840101     & XTI       & 2.07  \\ 
              &  5202840102     & XTI      &  3.46 \\ 
              &  5202840103     & XTI      &  9.00 \\ 
              &  5202840104     & XTI      &  5.35 \\ 
              &  5202840105     & XTI       & 2.43  \\ 
              &  5202840106     & XTI       &  0.23 \\ 
\noalign{\smallskip}  
\hline  
\noalign{\smallskip}  
\multicolumn{4}{c}{2025 outburst} \\
\noalign{\smallskip}  
\hline 
\noalign{\smallskip}  
\nustar               & 90501329001   & FPMA/FPMB        & 46.47 \\ 
\noalign{\smallskip}  
\nicer                      &  8202840101     & XTI      & 0.91   \\ 
                            &  8202840102     & XTI       & 1.12  \\ 
                            &  8202840103     & XTI       & 0  \\ 
                            &  8202840104     & XTI       & 0.22  \\ 
                            &  8202840105     & XTI       & 0.10  \\ 
                            \noalign{\smallskip}

\hxmt         &   P0704848001   & LE/ME/HE         &  3.43/6.52/3.76  \\
              
\noalign{\smallskip}  
\ep           &   01709175213   &  A(PW)/B(FF)     & 4.10  \\
              &   06800000586   &  A(TM)/B(TM)     & 3.02  \\
              &   06800000587   &  A(TM)/B(PW)     & 3.01  \\
              &   06800000589   &  A(TM)/B(PW)     & 1.78  \\
              &   06800000595   &  A(TM)/B(PW)     & 3.00  \\
              &   06800000597   &  A(PW)/B(PW)     & 4.27 \\

\noalign{\smallskip}  
\hline  
\end{tabular}  
}

\end{table} 

\subsection{\epfull}\label{sec:ep}

\ep\ was launched on January 9, 2024, equipped with WXT and FXT \citep{EP2025}. \ep\ has the capability to detect transients and perform rapid follow-up observations with excellent timing and spectral resolutions. FXT, including FXT-A and FXT-B, can operate in timing mode (TM), full frame  (FF) mode, and partial window (PW) mode \citep{EP-FXT}. The timing resolution of TM is 23.68 $\mu s$ \citep{EP-FXT-Timing}, which has been verified by the joint observations with \nicer\ for the AMXP SRGA J144459.2$-$604207 \citep{Li25}. 

On May 6, 2025, \ep/WFT reported an outburst from \psrtar.  \ep/FXT promptly triggered the target-of-opportunity (ToO) observations \citep{GCN40375}.  Six observations have been carried out between MJD UTC 60801.88--60807.26 (see Table~\ref{table:x-ray-observations}). We processed \ep/FXT observations using the tool \texttt{fxtchain}.  The source light curves, spectra, ancillary response files (ARFs), and 
response matrix files (RMFs)  were extracted from a circle region centered on the source position with a radius of $100''$, while the background spectra were from a nearby source-free region with the same radius. For the TM data we barycentered the event times using the tool \texttt{fxtbary}, adopting the JPL DE421 Solar System ephemeris and the source coordinates as listed in Table~\ref{table:eph}.

\subsection{\nicer}\label{sec:nicer}
\nicer\ \citep{nicer} observed \psrtar\ with the Obs. ID. 5202840101--5202840106 and 8202840101--8202840105 during its 2022 and 2025 outbursts, respectively. The results of the 2022 outburst observations have been reported by \citet{Sanna22}.  To perform a comprehensive coherent timing study, we analyzed all available \nicer\ observations from both outbursts. The data were processed with HEASOFT version 6.33 and the \nicer\ software NICERDAS version v12. We focused on the data from the calibrated unfiltered (UFA) event files. To identify and reject the time intervals of high background, we first extracted light curves in the 12--15~keV band, where the effective area for X-ray photons is negligible and events are dominated by charged particle interactions. Time intervals where this rate exceeded 5~cnt~s$^{-1}$ were identified as particle flares and excluded. Subsequently, we applied a count rate screening in the energy band of interest, retaining only time intervals with count rates less than 10~cnt~s$^{-1}$.
The UFA events were barycenter-corrected using the \texttt{barycor} tool and the JPL DE421 Solar System ephemeris.

\subsection{\hxmt}\label{sec:hxmt}
During the 2025 outburst, we triggered the \hxmt\ \citep{hxmt,hxmt-me,hxmt-he} ToO observations during MJD  60805.08--60805.48. The data were processed with the tool \texttt{hpipeline}, and the filtered exposure times for the low-energy \citep[LE;][]{hxmt-le}, middle-energy \citep[ME;][]{hxmt-me}, and high-energy \citep[HE;][]{hxmt-he} telescopes are presented in Table~\ref{table:x-ray-observations}. The events of ME and HE, generated from \texttt{mescreen} and \texttt{hescreen}, respectively, were barycentered using \texttt{hxbary}. 

\begin{figure*} %
\centering
\includegraphics[width=0.45\textwidth]{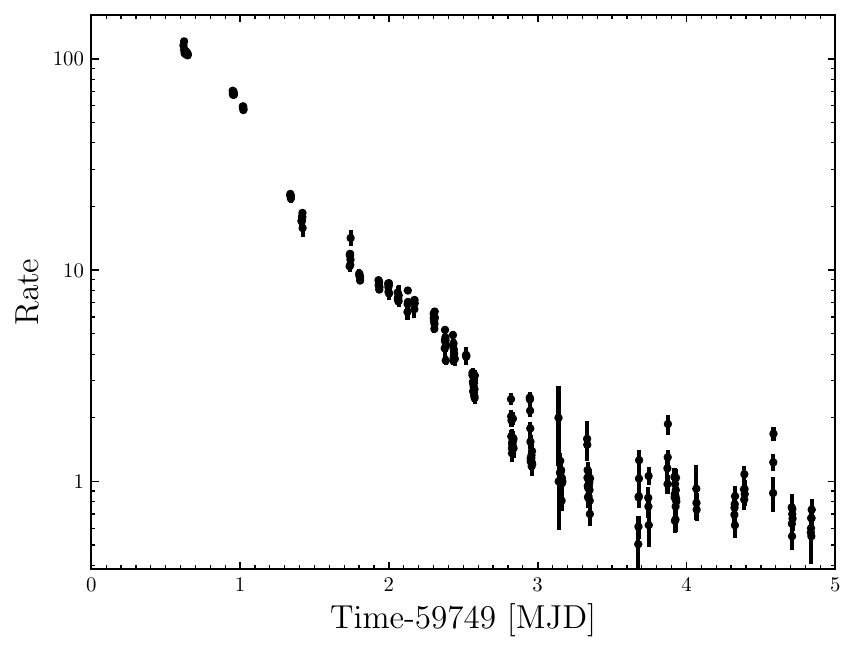}
\includegraphics[width=0.45\textwidth]{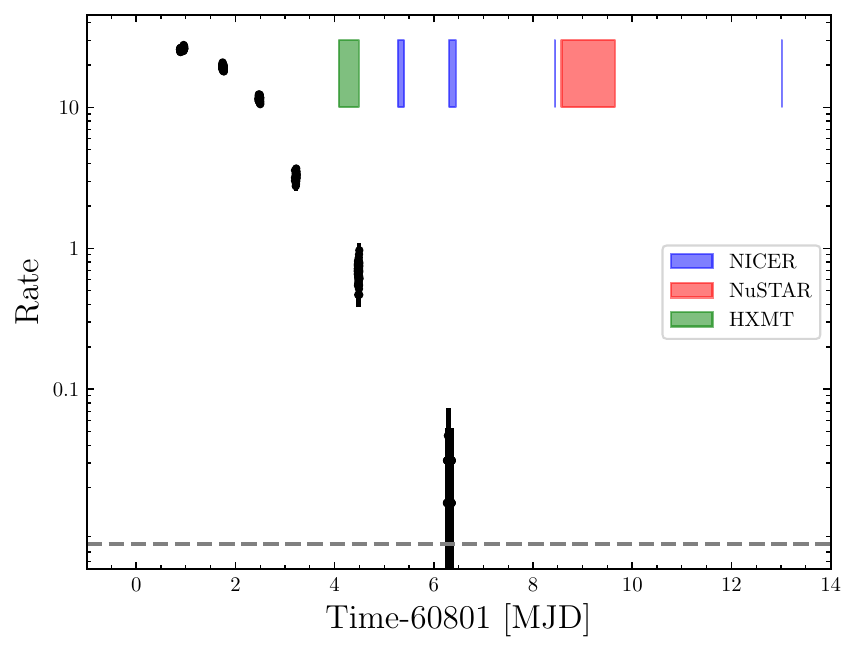}

\caption{Light curves of \psrtar\ from \nicer\ (left panel, 0.5--10 keV), and \ep\ (right panel, 0.5--10 keV) during its 2022 and 2025 outbursts, respectively. The vertical axes display the total count rate (source plus background) in units of counts per second on a logarithmic scale. In the right panel, the shaded rectangles mark the observational intervals of \nicer\ (blue), \nustar\ (red), and \hxmt\ (green). The dashed horizontal line indicates the background level for the \ep\ observations.
 }
\label{fig:outburst}
\end{figure*}

\subsection{\nustar}\label{sec:nustar}
We also analyzed an archival \nustar\ \citep{nustar} observation (Obs. ID 90501329001) of \psrtar\ taken during the 2025 outburst. The observation was performed over the MJD range 60809.57--60810.64, resulting in a total exposure of 46.47~ks. Data from both the FPMA and FPMB were processed separately using the standard \nustar\ pipeline tool \texttt{nupipeline}. However, no significant emission from \psrtar\ was detected in the data from either module. Consequently, these \nustar\ data are not considered further in our subsequent analysis.

\subsection{The outburst light curve}
The 2025 outburst light curve of \psrtar\ (\ep/FXT based) is presented in the right panel of Fig.~\ref{fig:outburst} and for comparison the 2022 outburst light curve (\nicer\, based) is shown in the left panel of Fig.~\ref{fig:outburst}. The right panel of Fig.~\ref{fig:outburst} also shows the observation windows of various other X-ray missions. The \ep/FXT observations began around MJD 60802.5, with a count rate of approximately 28 $\rm{cnt~s^{-1}}$.  While the rise of the outburst was not covered by these data, the subsequent evolution was characterized by a rapid and steep decay, with the source returning to a near-quiescent state by MJD 60806. Similar features have also been shown in the 2022 outburst (see the left panel in Fig.~\ref{fig:outburst}). The $\sim$4-day duration of this decay phase is consistent with the short-duration outburst behavior that is typical for this source.  The \hxmt, \nicer\ , and \nustar\ observations were carried out during the very end of the outburst decay phase or even during the subsequent quiescent state.  No type I X-ray bursts were found during these two outbursts.

\subsection{FAST}\label{sec:fast}
In 2024, we conducted four FAST observations of MAXI J1957+032, with a total duration of 1.67 hr. The observations were made using the 19-beam L-band receiver of FAST in the tracking mode, covering the 1.05--1.45\,GHz frequency band. The data were recorded in the PSRFITS format~\citep{Hotan2004}, with 8-bit resolution, 4096 frequency channels, four polarizations, and a sampling time of 49.152\,$\mu$s \citep{Jiang2020}.

We performed a pulsar search using \texttt{PRESTO}\footnote{\url{https://github.com/scottransom/presto}}~\citep{Ransom2001}. The optimal dedispersion strategies were determined using the script \texttt{DDplan.py}, covering dispersion measure up to 1000 ${ \rm pc ~ cm}^{-3}$, and applied the acceleration search with an integration time ($t_{\rm int} \sim ~$6 min) spanning 10\% of the orbital period. The \texttt{accelsearch} routine was employed with parameter $z_{\text{max}} = 200$, which represents the maximum number of Fourier bins that the highest harmonic can drift linearly in the power spectrum. No significant radio pulsations were detected. The upper limit on the source flux density, i.e., the minimum detectable flux density, $S_{\rm min}$, is given by \citep{van2010}

\begin{equation}
\label{eq1}
S_{\rm min} = \frac{(S/N)\beta T_{\rm sys}}{G(n_{\rm p}t_{\rm int}\Delta F)^{1/2}} \left(\frac{W}{P-W}\right)^{1/2},
\end{equation}
where $S/N$ is the threshold in the signal-to-noise ratio for the pulsar detection, $n_{\rm p}$ is the number of polarizations, $\beta$ is the sensitivity degradation factor, $T_{\rm int}$ is the integration time, $\Delta F$ is the observing frequency bandwidth, $P$ and $W$ are the pulsar spin period and pulse width, respectively, $G$ is the telescope gain, and $T_{\rm sys}$ is the system temperature. For our observations, the $G$ and $T_{\rm sys}$ values, which vary with the zenith angle, are listed in Table~\ref{table:observation_fast}. We used $\beta = 1$, $S/N = 7$, $n_{\rm p} = 2$, $\Delta F = 400 ~\rm {MHz}$, and $W/P = 0.3$. Adopting these parameters, the $7\sigma$ flux density upper limit for each observation was constrained to $12.3\sim 16.3 \,\mu\rm{Jy}$ (see Table~\ref{table:observation_fast}).

\begin{table}%
{\small
\caption{Details of FAST observations and data reduction. \label{table:observation_fast} }

\centering
\begin{tabular}{lccccc} 
\hline 
Obs. No.  & Obs. MJD  & Length & $G$ & $T_{\rm sys}$ &Sensitivity \\
 &    & (min) & $(\rm{K~Jy}^{-1})$ & (K) & $(\mu\text{Jy})$ \\
\hline 
01 &  60335.25  & 30 & 14.6 & 29.3 & 16.3 \\  
02 &  60365.10  & 30 & 16.2 & 24.4 & 12.3 \\ 
03 &  60394.98  & 20 & 16.1 & 27.6 & 14.0 \\ 
04 &  60424.91  & 20 & 16.2 & 26.3 & 13.2 \\ 

\hline  
\end{tabular}  
}

\end{table} 

\begin{table}[t] 
{\small
\caption{Orbital and spin parameters of \psrtar\ derived in this work from a 4D optimization scheme using \nicer\ 0.5--10 keV data for the 2022 outburst and  \ep\ 0.5--10 keV data for the 2025 outburst. }
\centering
\begin{tabular}{lc} 
\hline \hline 
Parameter                      & Values                               \\
\hline 
\noalign{\smallskip}  
$\alpha_{2000}$ $^{a}$               & $19^{\hbox{\scriptsize h}} 56^{\hbox{\scriptsize m}} 39\fs11$       \\   
$\delta_{2000}$ $^{a}$               & $03\degr26\arcmin43\farcs7$       \\        \noalign{\smallskip}       
$ e $                          & $0$ (fixed)                      \\             
\hline
\noalign{\smallskip}  
\multicolumn{2}{c}{2025 outburst: \ep}\\
\noalign{\smallskip}  
\hline 
$ P_{\rm orb} $ (s)                &3652.95(43)       \\             
$ a_{\rm x}\sin i$ (lt-s)            & 0.013\,85(12)                       \\             
$T_{\rm asc} $ (MJD, TDB)                & 60802.746\,344(60)             \\    
\noalign{\smallskip}  

\noalign{\smallskip} 
Validity range  (MJD, TDB)                & 60802 -- 60807          \\              
$t_0$ (Epoch, MJD, TDB)                  &  60804                       \\
$\nu$ (Hz)                          & 313.643\,736\,8(7)             \\
JPL Ephemeris                  & DE421                             \\   
\hline
\noalign{\smallskip}  
\multicolumn{2}{c}{2022 outburst: \nicer}\\
\noalign{\smallskip}  
\hline 
$ P_{\rm orb} $ (s)               & 3652.94(15)                   \\             
$ a_{\rm x}\sin i$ (lt-s)             & 0.013\,80(7)                \\             
$T_{\rm asc} $ (MJD, TDB)                 & 59749.633\,151(43)            \\             
\noalign{\smallskip}  

\noalign{\smallskip}  
Validity range (MJD, TDB)                &  59749 -- 59755   \\                      
$t_0$ (Epoch, MJD, TDB)                  &  59752.0                      \\
$\nu$ (Hz)                         & 313.643\,741\,98(24)           \\
JPL Ephemeris                  & DE421                             \\         

\hline 

\end{tabular}
\label{table:eph} 
    \tablefoot{$^{a}$The position used in the barycentering process \citep{ATel9591} .}
}
\end{table} 

\section{Timing results}\label{sec:timing}

\subsection{The 2022 outburst}

We reanalyzed the \nicer\ data from the 2022 outburst of \psrtar, originally presented by \citet{Sanna22}. Their timing solution provided an orbital and spin ephemeris for the source, yielding a spin frequency, $\nu$, of $313.64374049(22)$~Hz. However, they introduced a phase jump of $\sim0.2$ around MJD 59750.2, increasing the number of free parameters from 5 to 7.

To investigate the timing solution for this dataset we employed a 4D optimization scheme based on a downhill \texttt{SIMPLEX} algorithm (see e.g., \citealt{deFalcoa,ZLi21,Li23} for the earlier
3D version, and \citealt{Li24,Li25} for the current 4D version of the method). This method was designed to efficiently locate the maximum of the $Z^2_3$ test statistic by simultaneously optimizing the spin frequency, $\nu$, the orbital period, $P_{\rm orb}$, the projected semimajor axis, $a_x \sin i$, and the time of the ascending node, $T_{\rm asc}$ (assuming a circular orbit; four free parameters). We used the parameters from \citet{Sanna22} as initial values for our optimization routine.

Our analysis resulted in an alternative, equivalent timing solution, with the parameters presented in Table~\ref{table:eph}. Our derived values for $P_{\rm orb}$, $a_x \sin i$, and $T_{\rm asc}$ are all consistent with the results of \citet{Sanna22} at the 1$\sigma$ confidence level. However, we measure a spin frequency of $\nu = 313.64374198(24)$~Hz, which differs from \citet{Sanna22} by approximately 6$\sigma$. Alternatively, if we searched for the optimum spin frequency of the first \nicer observation and the remaining five observations separately, i.e., before and after their introduced phase jump, we obtained for both sets the same value for the spin frequency as is reported in \citet{Sanna22}.

\subsection{The 2025 outburst}

We analyzed observations from multiple X-ray missions during the 2025 outburst, including \nicer, \nustar, \hxmt, and \ep. We performed separate searches for coherent pulsations in the datasets from each of the four instruments. For each dataset, we employed a method similar to that applied to the 2022 outburst data, using a downhill \texttt{SIMPLEX} algorithm to optimize the spin and orbital parameters by maximizing the $Z^2_3$ statistic.

No significant pulsations were detected from \psrtar\ in the \nicer, \nustar, or \hxmt\ observations. This can be explained because their observation windows miss the ON state of the outburst (see right panel of Fig. \ref{fig:outburst}). In contrast, \ep\ successfully monitored the source during the main activity of the outburst. We
performed a timing analysis of the \ep\ data taken in TM during
MJD 60802.73--60805.47 with a total exposure time of 13.79 ks. Our independent search of the \ep\ data yielded a strong detection with a $Z^2_3$ statistic value of approximately 197 ($13.2 \sigma$). The resulting best solution for the 2025 outburst is presented in Table~\ref{table:eph}.

We compared these new parameters to those from our globally optimized solution for the 2022 \nicer\ data. The orbital period, $P_{\rm orb}$, and the projected semimajor axis, $a_x \sin i$, from the 2025 outburst are fully consistent with our results from the 2022 outburst within their 1$\sigma$ uncertainties. The newly measured time of the ascending node, $T_{\rm asc}$, of $60802.746344(60)$ and spin frequency, $\nu$, of $313.6437368(7)$~Hz at a reference epoch, $t_0$, of $60802.0$~MJD -- separated from the 2022 observations by approximately three years -- provide a long baseline that is useful for studying the long-term evolution of the pulsar's spin frequency (see Sect.~\ref{sec:spin_down}).

\subsection{The pulse profile morphology}
\begin{figure}[t]
\centering
\includegraphics[width=9cm]{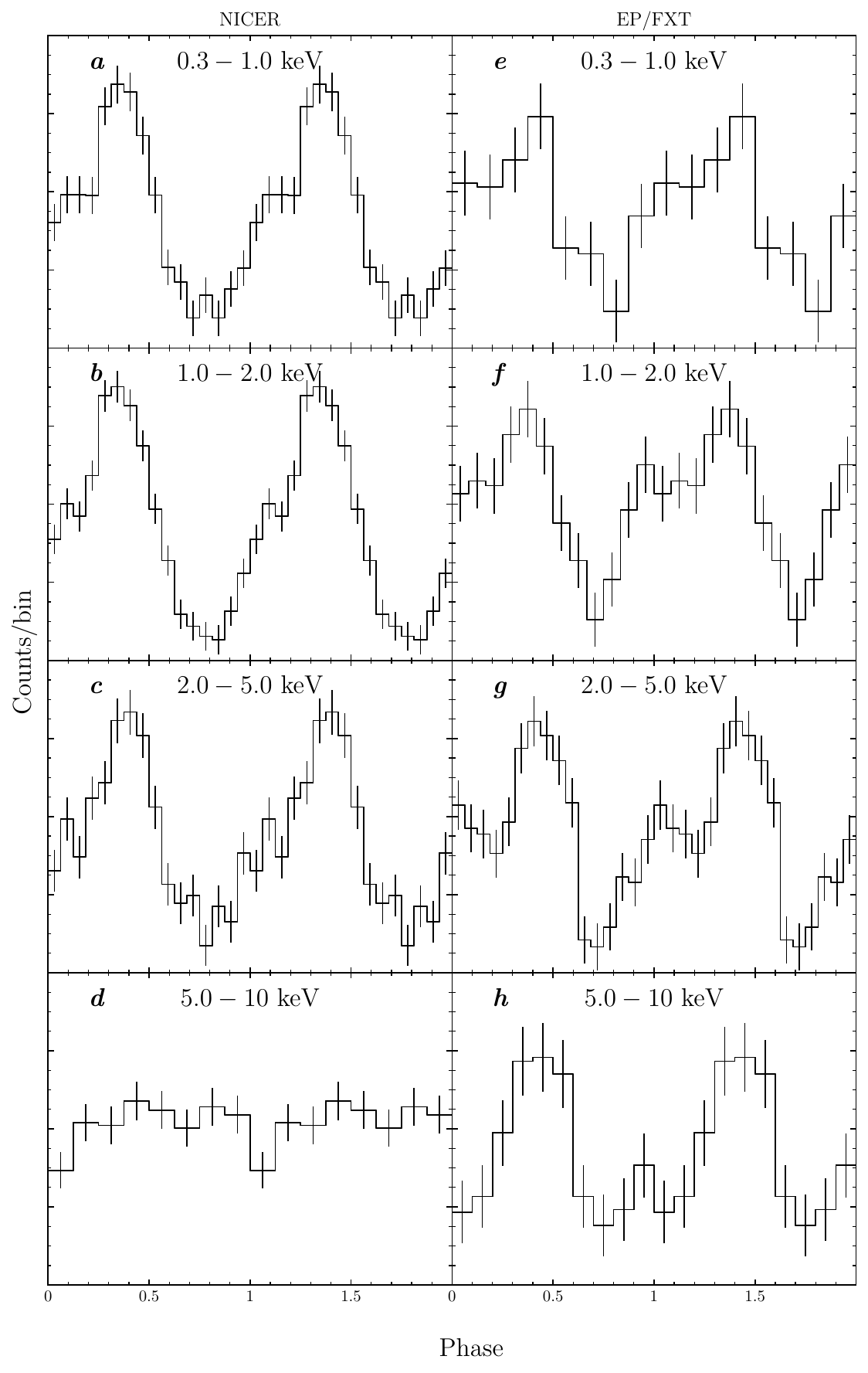}
\caption{Energy-resolved pulse profiles of \psrtar\ during its 2022 outburst (left panels, $\textbf{\textit{a}}$ -- $\textbf{\textit{d}}$; $0.3-10$ keV, observed with NICER) and 2025 outburst (right panels, $\textbf{\textit{e}}$ -- $\textbf{\textit{h}}$; $0.3-10$ keV, observed with EP/FXT). Above 5 keV, the pulsed emission was undetectable in the 2022 outburst, but significant in the 2025 outburst. We show two cycles for clarity, while the error bars represent $1\sigma$ errors.   }
\label{fig:nicer_pulse_profile}
\end{figure}

Using the orbital parameters listed in Table~\ref{table:eph}, we corrected the barycentered event times from \nicer\ and \ep\ observations for the periodic orbital motion effects.  For the 2022 outburst, we obtained after pulse-phase folding pulse-phase distributions (pulse-profiles) for the following energy ranges: 0.3--1.0, 1.0--2.0, 2.0--5.0, and 5.0--10 keV (see the left panels in Fig.~\ref{fig:nicer_pulse_profile}). No significant pulsed emission was detected above 5 keV. The folded profile in this band is consistent with Poisson noise, and we place a $3\sigma$ upper limit on the pulsed amplitude of 4.6\%.

For the 2025 outburst, we folded the EP timing data in the same energy ranges as the 2022 outburst (see the right panels of Fig.~\ref{fig:nicer_pulse_profile}). The pulsations have now been detected across the whole energy band up to $\sim 10$ keV, indicating that the pulsed emission is harder than the 2022 outburst emission. After phase-aligning the profiles from both epochs by simple cross-correlation for a direct comparison, we found that the overall pulse shape is remarkably similar.  The profiles from both the 2022 and 2025 outbursts were quite comparable, each showing a main peak at a phase of $\sim0.4$ and a secondary peak at a phase of $\sim0$. If the pulsed emission is produced from hot spot(s) at the magnetic pole(s), it suggests that the geometrical configuration between the magnetic pole, spin axis, and observer's viewing angle has not changed significantly between 2022 and 2025.

\begin{figure}[t]
\centering
\includegraphics[width=9cm]{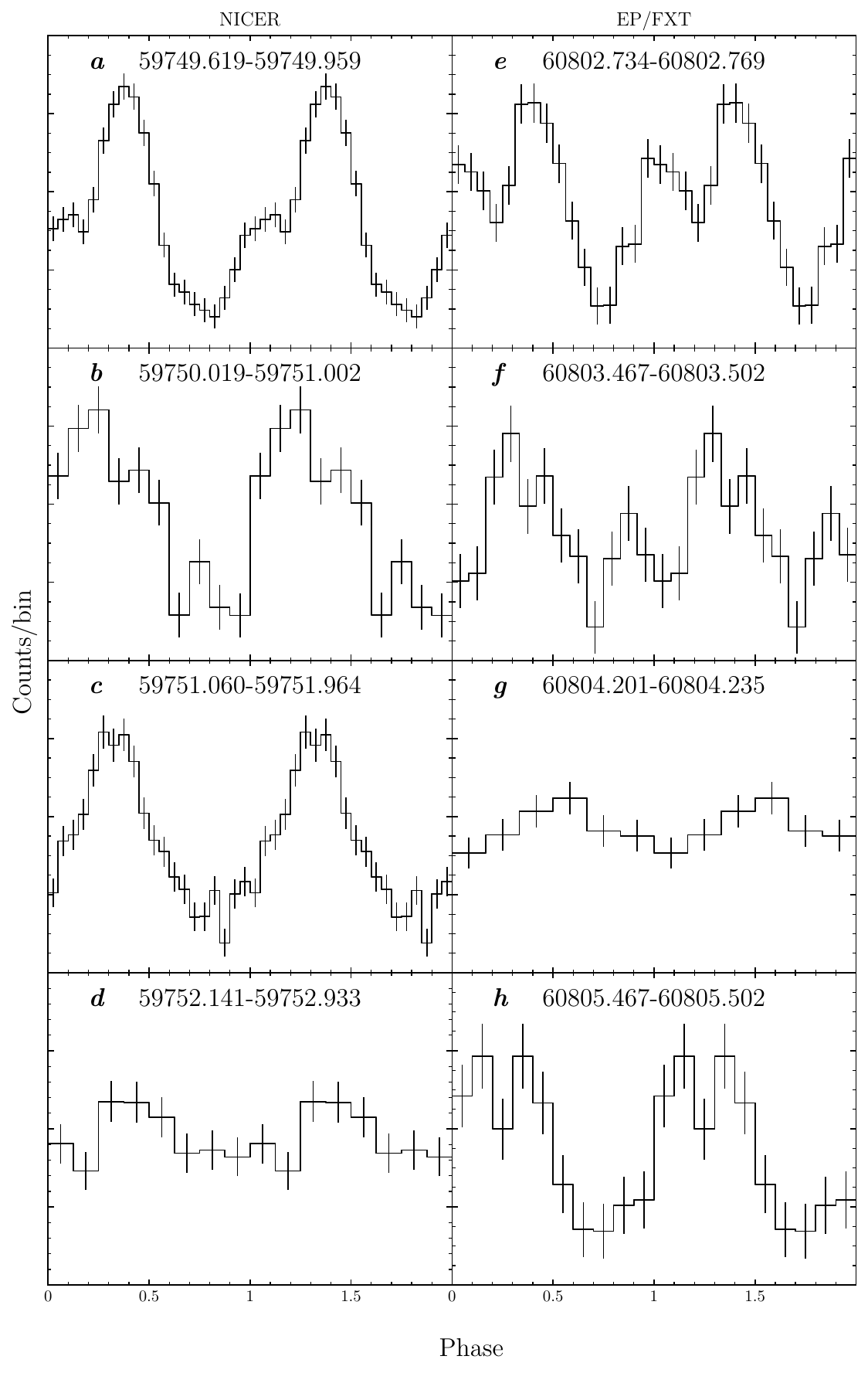}
\caption{Time-resolved pulse profiles of MAXI J1957+032 during its 2022 outburst (left panels, $\textbf{\textit{a}}$ -- $\textbf{\textit{d}}$; $0.3-5$ keV, observed with NICER) and its 2025 outburst (right panels, $\textbf{\textit{e}}$ -- $\textbf{\textit{h}}$; $0.3-10$ keV, observed with EP/FXT).  We show two cycles for clarity, while the error bars represent $1\sigma$ errors.  The time intervals in MJD are shown in the middle of the panels. }
\label{fig:profile_time_resolved}
\end{figure}

We also obtained time-resolved pulse profiles for the 2022 and 2025 outbursts to study the evolution of the pulse morphology. The time intervals for these profiles were defined based on the segmentation of the individual observation blocks. Specifically, the four panels on the left side of Fig.~\ref{fig:profile_time_resolved} ($\textbf{\textit{a}}$--$\textbf{\textit{d}}$) correspond directly to the first four distinct \nicer\ observations during the 2022 outburst. Similarly, for the 2025 outburst observed by \ep, the four panels on the right ($\textbf{\textit{e}}$--$\textbf{\textit{h}}$) correspond directly to the observations 06800000586, 06800000587, 06800000589, and 06800000595, respectively, where TM mode data were available.

No significant pulsations have been detected in the time intervals of panels ($\textbf{\textit{d}}$) and ($\textbf{\textit{g}}$) in Fig.~\ref{fig:profile_time_resolved}.  Notably, for the interval corresponding to panel ($\textbf{\textit{g}}$) (Obs. 3; MJD 60804.201--60804.235), the detection significance was only $2.5\sigma$. This non-detection is intriguing because the total count rate during this observation was approximately 2.5 times higher than in the subsequent \ep\ observation (Obs. 4; panel $\textbf{\textit{h}}$), where pulsations were significantly detected ($5.7\sigma$). This behavior is likely indicative of pulse intermittency, a phenomenon observed in other AMXPs such as IGR J17498--2921 \citep{Li24}, SAX J1748.9--2021 \citep{Altamirano08}, and HETE J1900.1--2455 \citep{Galloway07}, possibly in combination with an intrinsic evolution of the pulsed fraction as the outburst decays \citep[see e.g.,][ Fig. 1 for the 2022 outburst]{Sanna22}.

For the 2022 outburst observed by NICER, the profile initially exhibits a main peak accompanied by a shoulder in panel ($\textbf{\textit{a}}$), which then transited into a more defined single-peaked structure in panels ($\textbf{\textit{b}}$)-($\textbf{\textit{c}}$). A similar evolutionary pattern was observed for the 2025 data from EP/FXT, where the initial profiles in panels ($\textbf{\textit{e}}$) and ($\textbf{\textit{f}}$) also showed a main peak with a shoulder, before evolving into a single-peaked profile in panel ($\textbf{\textit{h}}$). Moreover, beyond the changes in shape within each outburst, the phase of the main peak also clearly shifted during these two outbursts. This significant and complex evolution of the pulse morphology prevents us from performing a reliable time-of-arrival (TOA) analysis as has been done for other AMXPs that exhibit more stable pulse profiles \citep[see e.g.,][]{Li23,Li24,Li25}.

\section{Spectral analysis}\label{sec:spec}

We performed a spectral analysis of the 2025 outburst using the data collected by EP/FXT. For our analysis, we prioritized data from the PW mode when available. For the observation for which PW mode data was not available (Obs. ID 06800000586), we used data from the TM. The spectra from individual observations were extracted, and fit in the 0.5--10~keV range using \texttt{XSPEC} version 12.14.1 \citep{arnaud96}, adopting the \texttt{tbabs*powerlaw} model with abundances from \citet{wilms00}. The free parameters are the hydrogen column density for \texttt{tbabs}, and the power-law index, $\Gamma$, and normalization.

\begin{figure}[t]
\centering
\includegraphics[width=9cm]{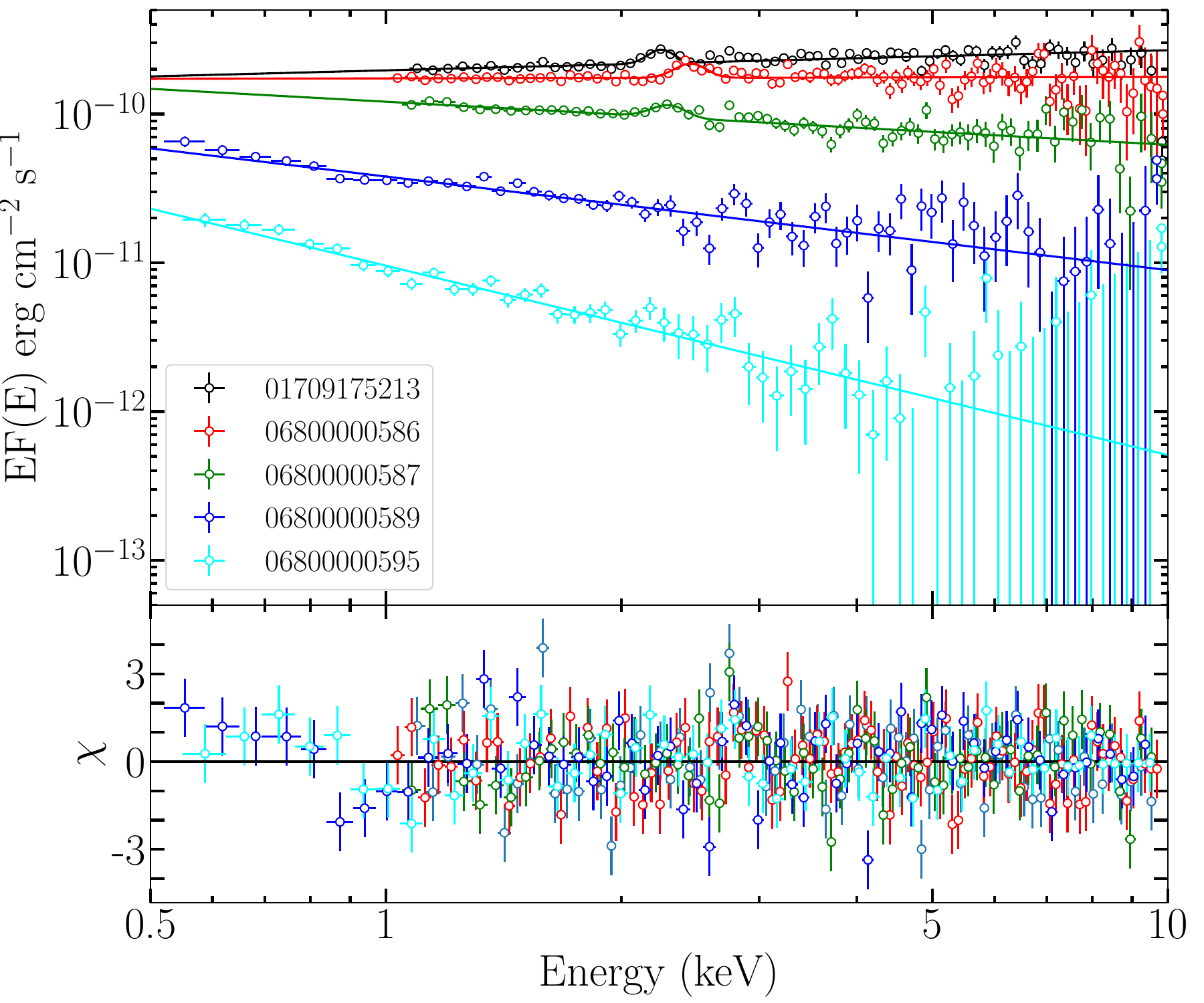}
\caption{Unfolded spectra and the best-fit models of the first five EP observations. For the first three observations, an instrumental Gaussian component is needed. Moreover, these spectra show strong excess below 1 keV, and are thus ignored.   }
\label{fig:spectra}
\end{figure}

\begin{figure}[t]
\centering
\includegraphics[width=9cm]{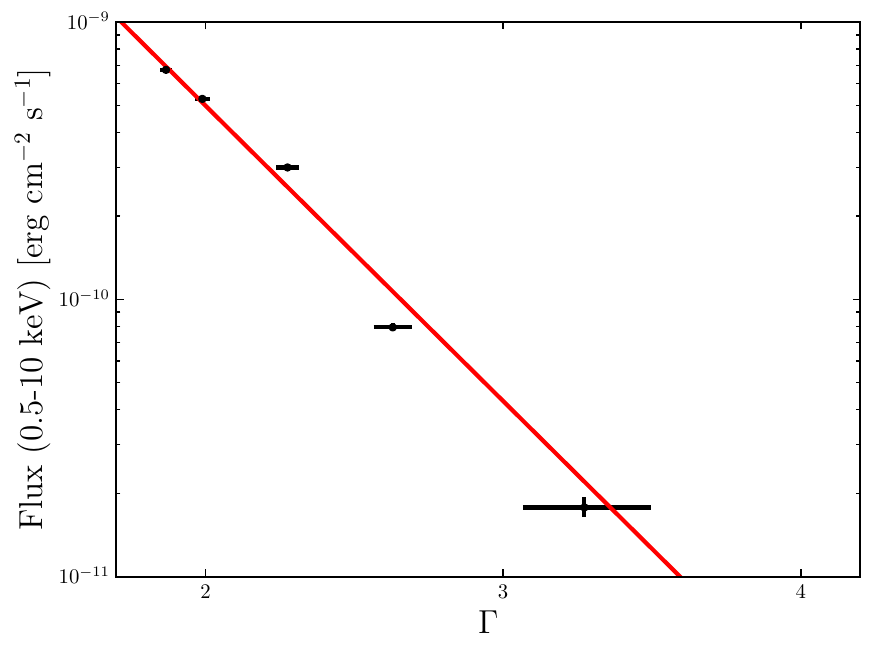}
\caption{Relation between the unabsorbed flux in 0.5--10 keV and the power law index, $\Gamma$. The red line represents the best fit, $F_X\sim\Gamma^{-(2.5\pm0.3)}$. }
\label{fig:flux_gamma}
\end{figure}

This model provided an acceptable description of the spectra, with the best-fit reduced $\chi^2$,  $\chi^2_{\nu}\sim0.9-1.6$. The parameters for the best fit are listed in Table~\ref{tab:best-parameter}. No significant trends are visible in the residuals (see Fig.~\ref{fig:spectra}).  The analysis reveals a clear and significant spectral evolution throughout the outburst decay. The hydrogen column density, $N_{\rm H}$, remained low with best-fit values in the range of $(1 - 3) \times 10^{21}~{\rm cm^{-2}}$. In the initial observation, the source was brightest with an unabsorbed 0.5--10~keV flux of $\sim 6.7 \times 10^{-10}~{\rm erg~s^{-1}~cm^{-2}}$, and the spectrum was correspondingly hard, with a power-law photon index of $\Gamma \sim 1.87$. Over the subsequent days, the flux decayed by a factor of more than 30. This decay was accompanied by a systematic softening of the spectrum, with the photon index steepening to a final measured value of $\Gamma \sim 3.27$ in the last significant detection, which had a flux of $\sim 1.8 \times 10^{-11}~{\rm erg~s^{-1}~cm^{-2}}$, see Fig.~\ref{fig:spectra}. Moreover, this correlation can be well described as $F_X\sim\Gamma^{-(2.5\pm0.3)}$; see Fig.~\ref{fig:flux_gamma}.  This trend of spectral softening with decreasing luminosity  is consistent with the behavior observed in previous outbursts of this source \citep{MataSanchez2017} and matches the evolution reported by \citet{Sanna2025} based on \swift\ monitoring of the 2025 outburst.  Similar spectral evolution has also been reported for other AMXPs such as  SAX J1808.4--3658 \citep{Campana08}.

We note that during the first three observations, when the source was brightest, a narrow residual feature was consistently observed around 2.4~keV. This feature was well modeled by including a Gaussian component in the model. We attribute this feature to an instrumental origin, likely related to calibration uncertainties. Moreover, these spectra showed some excess below 1 keV,  a feature that was notably absent in the 2022 dataset \citep{Sanna22}, which was ignored in our analysis. The final observation was consistent with a non-detection and yielded only a weak constraint on the spectral parameters.

\begin{table*}
	\centering
	\caption{Best-fit spectral parameters for different \ep observations performed during the 2025 outburst of MAXI J1957+032. }
	\label{tab:best-parameter}
	\begin{tabular}{lcccc}
		\hline
		\hline
		Obs. ID & $N_{\rm H}$ & $\Gamma$ & Unabsorbed Flux & $\chi^2$/d.o.f. \\
		& ($10^{22}$ cm$^{-2}$) & & ($10^{-10}$ erg cm$^{-2}$ s$^{-1}$) & ($\chi^2_\nu$) \\
		\hline
		01709175213 & $0.14 \pm 0.01$ & $1.87 \pm 0.02$ & $6.73 \pm 0.05$ & 119.4/74 (1.61) \\
		06800000586 & $0.22 \pm 0.02$ & $1.99 \pm 0.03$ & $5.29 \pm 0.06$ & 97.7/96 (1.02) \\
		06800000587 & $0.31 \pm 0.03$ & $2.28 \pm 0.04$ & $2.99 \pm 0.07$ & 78.8/68 (1.16) \\
		06800000589 & $0.31 \pm 0.02$ & $2.63 \pm 0.07$ & $0.79 \pm 0.03$ & 97.6/66 (1.48) \\
		06800000595 & $0.34 \pm 0.05$   & $3.27\pm0.22$ & $0.18 \pm 0.02$ & 52.8/60 (0.88) \\
		\hline
	\end{tabular}
    \tablefoot{The spectra were fit with the \texttt{tbabs*powerlaw} model for the 0.5--10 keV range. The unabsorbed fluxes in 0.5--10 keV are also reported. Errors are quoted at the $1\sigma$ confidence level.}
\end{table*}

\section{Discussion}\label{sec:dis}
In this work we have presented a comprehensive multiwavelength analysis of the AMXP \psrtar, covering its major outbursts in 2022 and 2025, as well as the intervening quiescent period. We have established high-precision timing solutions for these two distinct epochs, characterized the spectral evolution during the 2025 outburst, and placed a tight constraint on the source's radio emission during X-ray quiescence.

For the 2022 outburst, our reanalysis of the \nicer\ data using a robust 4D optimization algorithm yielded a globally optimized timing solution. This newly derived ephemeris provided an independent and comparable solution to the one obtained by \citet{Sanna22} without the introduction of any phase jump at a certain epoch (i.e., adding two more fit parameters). For the 2025 outburst, we reported a pulsation detection of $13.2\sigma$ from this source with \ep. The analysis of this unique dataset provided a new timing solution for a second epoch, which was crucial as other X-ray missions did not detect pulsations during this event. Furthermore, our analysis of the EP/FXT spectra from the 2025 outburst revealed a clear spectral evolution, with the power-law photon index softening from $\Gamma \sim 1.9$ to $\Gamma \sim 3.3$ as the outburst decayed, a behavior that is characteristic of AMXPs. The successful acquisition of the 2025 outburst timing parameters is the key result that enables the investigation of the long-term spin behavior of the system.

It is worth noting that while this manuscript was under review, an independent study by \citet{Sanna2025} appeared, reporting on the 2025 outburst using data from \xmm. A comparison of the timing solutions reveals that the orbital parameters derived from \xmm\ are highly consistent with our \ep\ results within $1\sigma$, and the measured spin frequencies are consistent within $\sim 2\sigma$.

\subsection{Long-term spin-down rate}\label{sec:spin_down}

By combining the spin frequency measured during the 2022 outburst, $\nu_{2022} = 313.643~741~98(24)$~Hz, with that of the 2025 outburst, $\nu_{2025} = 313.643~736~8(7)$~Hz, we can constrain the long-term spin evolution of the pulsar. We define a quiescent baseline from the end of the 2022 outburst (MJD 59755) to the beginning of the 2025 outburst (MJD 60802). Assuming a constant spin-down during this interval, we calculated an average spin frequency derivative of $\dot\nu = (-5.73 \pm 0.28) \times 10^{-14}~{\rm Hz~s^{-1}}$.   If this spin-down rate during the quiescent state is caused by (rotating) magnetic dipole emission, the magnetic dipole moment is \citep{Spitkovsky06,Riggio11} 
\begin{equation}\label{equ:mag2}
\mu_{26}=26.15\times (1+\sin^2\theta)^{-1/2}I_{45}^{1/2}\nu_2^{-3/2}(-\dot\nu_{-14})^{1/2},
\end{equation}
where $\theta$ is the angle between the rotation and magnetic axes, $I_{45}$ is the NS's moment of inertia in units of $10^{45}~{\rm g~cm^2}$, the spin frequency, $\nu_2=\nu/100$ Hz, and the spin frequency derivative, $\dot\nu_{-14}=\dot\nu/10^{-14}~{\rm Hz~s^{-1}}$. We adopted a NS radius of 11 km, $\sin^2 \theta\sim0-1$, and $I_{45}=1.5$, and obtained a magnetic dipole moment of $(9.8-13.4)\times10^{26}~{\rm G~cm^3}$, corresponding to a relatively high surface magnetic field of  $(7.3-10.4)\times10^8~\rm{G}$. Alternatively, if we adopted the spin frequency of $\nu_{2022} = 313.643~740~49(22)$~Hz that is reported in \citet{Sanna22}, we obtain a spin-down rate of $\dot\nu = (-4.08 \pm 0.81) \times 10^{-14}~{\rm Hz~s^{-1}}$ and magnetic field of $(6.2-8.7)\times10^8~\rm{G}$. Both values are slightly smaller than the aforementioned ones.

Recently, \citet{Sanna2025} derived a smaller spin-down rate of $\dot{\nu} \approx -2.25 \times 10^{-14}$~Hz~s$^{-1}$. While the choice of the 2022 baseline plays a role, the primary driver of this discrepancy is the difference in the 2025 spin frequency measurement: our \ep\ value is lower than their \xmm\ value ($\nu_{\rm XMM} = 313.643~738~44(34)$ Hz) by $\sim 1.6 ~\mu$Hz. It implies a larger frequency drop over the 3-year baseline, and thus a faster spin-down rate, regardless of which 2022 solution is used as a reference. However, despite these differences in $\dot{\nu}$, the inferred magnetic field strength derived by \citet{Sanna2025} ($B \lesssim 10^9$~G) remains broadly consistent with our estimate.

\subsection{The non-detection of radio pulsation}

The radio behavior of \psrtar\ during its X-ray quiescent phase was previously uncharacterized. One of the primary motivations for our work was to determine if this source activates as a radio pulsar once accretion halts, a fundamental question for any such system. The pulsar recycling scenario predicts that this transition should occur, but a robust theory of when or how is absent \citep[see e.g.,][]{Papitto22}. The empirical evidence is also ambiguous: a rare class of transitional millisecond pulsars (tMSPs), such as IGR J18245--2452, cleanly demonstrates this state-switching behavior \citep{papitto13c}. In contrast, the vast majority of the wider AMXP population has evaded detection in numerous deep radio pulsation searches, with notable non-detections including Aql X--1 \citep{Burgay03,Peng25}, XTE J0929--314 \citep{Iacolina09}, as well as XTE J1751--305, XTE J1814--338, and SAX J1808.4--3658 \citep{Iacolina10,Patruno17}. Our observations aimed to establish for the first time whether \psrtar\ behaves like a tMSP or like the majority of radio-quiet AMXPs.

Our deep search for radio pulsations from \psrtar\ using FAST yielded a non-detection, for which we establish a 7$\sigma$ upper limit on the flux density of 12.3 $\mu$Jy. Given the substantial spin-down power of $\dot E=-4\pi^2I\nu\dot\nu \approx 1.1 \times 10^{36}~{\rm erg~s^{-1}}$, the obtained upper limit of the flux density corresponds to a radio luminosity limit of $L_{\rm radio} < 2.87 \times 10^{26}~{\rm erg~s^{-1}}$  in 1.05--1.45 GHz at a maximum distance of 7 kpc \citep{Ravi17}. This implies a radio conversion efficiency of $\xi = L_{\rm radio} / \dot E < 2.8 \times 10^{-10}$. This efficiency is significantly lower than the typical range of $10^{-8}-10^{-5}$ observed for MSPs with comparable spin-down power \citep{Szary14}, indicating that the non-detection is intrinsic to the source, not a result of insufficient instrumental sensitivity.

Several scenarios can explain this quietness of radio pulsation \citep[see][and references therein]{Peng25}. The first is a purely geometric effect, whereby the pulsar's narrow radio beam never intersects our line of sight. While this is a valid consideration for any single object, it becomes a less tenable explanation when applied to the growing population of non-detections. In the future, as many more quiescent AMXPs are targeted by deep searches with highly sensitive instruments like FAST, we will be able to test this hypothesis statistically. If radio pulsations continue to be systematically absent across a large sample, the geometric beaming explanation can be ruled out.  Another compelling explanation is that a low-level accretion continues even during the X-ray quiescent phase. Even in quiescence, if the system operates in the propeller regime or undergoes radio ejection, a significant amount of material can be expelled from the inner system but remains in the vicinity. This ongoing inflow of matter, however faint, can effectively quench the radio emission by providing enough plasma to either absorb the radio signal via free-free absorption, or, more fundamentally, to suppress the accelerating electric potentials within the magnetospheric gaps required to power the coherent emission mechanism.

\begin{acknowledgements}
We appreciate the referee for constructive comments and suggestions, which improved the manuscript. 
This work was supported by the Major Science and Technology Program of Xinjiang Uygur Autonomous Region (No. 2022A03013-3) and China's  Space Origins Exploration Program. Z.S.L. and Y.Y.P. were supported by National Natural Science Foundation of China (12273030, 12103042), the science and technology
innovation Program of Hunan Province (No. 2024JC0001). This work is supported by the China Manned Space Program with grant no CMS-CSST-2025-A13. This work made use of data from the \hxmt\
mission, a project funded by China National Space Administration (CNSA) and the Chinese Academy of Sciences (CAS), and also from the High Energy Astrophysics Science Archive Research Center (HEASARC), provided by NASA’s Goddard Space Flight Center. EP
is a space mission supported by Strategic Priority Program on Space Science
of Chinese Academy of Sciences, in collaboration with ESA, MPE and
CNES. FAST is a Chinese national megascience facility,
operated by National Astronomical Observatories, Chinese
Academy of Sciences. The research is partly supported by the
Operation, Maintenance and Upgrading Fund for Astronomical
Telescopes and Facility Instruments, budgeted from the
Ministry of Finance of China (MOF) and administrated by
the Chinese Academy of Sciences (CAS).

\end{acknowledgements}

\bibliography{pulsars}{}
\bibliographystyle{aa}

\end{document}